\documentclass[
 reprint,
superscriptaddress,
 amsmath,amssymb,
 aps,
 prl,
]{revtex4-2}

\usepackage[english]{babel}
\usepackage[utf8]{inputenc}
\usepackage[colorinlistoftodos, color=green!40, prependcaption]{todonotes}
\usepackage{amsthm}
\usepackage{mathtools}
\usepackage{physics}
\usepackage{xcolor}
\usepackage{graphicx}
\usepackage[left=23mm,right=13mm,top=35mm,columnsep=15pt]{geometry} 
\usepackage{adjustbox}
\usepackage{placeins}
\usepackage[T1]{fontenc}
\usepackage{lipsum}
\usepackage{csquotes}
\usepackage[pdftex, pdftitle={Article}, pdfauthor={Author}]{hyperref} 
\usepackage{lineno}
\usepackage{multirow}

\begin{document}
\title{Constraining the trend of the $N = 50$ shell gap towards $^{100}$Sn with the masses of $^{96-98}$Cd}
\author{D. Lange}
    \email[Corresponding author: ]{daniel.lange@mpi-hd.mpg.de}
    \altaffiliation[\\This Letter contains data from the Ph.D. thesis work of Daniel Lange, enrolled at the Ruprecht-Karls-Universität Heidelberg.\\]{}
    \affiliation{Max-Planck-Institut für Kernphysik, 69117 Heidelberg, Germany}
    
\author{D. Atanasov}
    \affiliation{Studiecentrum voor Kernenergie SCK-CEN, 2400 Mol, Belgium}
\author{M. Au}
    \affiliation{European Organization for Nuclear Research (CERN), 1211 Geneva 23, Switzerland}
    \affiliation{Johannes Gutenberg-Universität Mainz, 55099 Mainz, Germany}
\author{A.~Belley}
    \affiliation{Massachusetts Institute of Technology, Cambridge, Massachusetts 02139, USA}
\author{M. Benhatchi}
    \affiliation{Université Paris-Saclay, CNRS/IN2P3, IJCLab, 91405 Orsay, France}
\author{K. Blaum}
    \affiliation{Max-Planck-Institut für Kernphysik, 69117 Heidelberg, Germany}
\author{R. B. Cakirli}
    \affiliation{Max-Planck-Institut für Kernphysik, 69117 Heidelberg, Germany}
\author{P. F. Giesel}
    \affiliation{Institut für Physik, Universität Greifswald, 17489 Greifswald, Germany}
\author{A. Herlert}
    \affiliation{FAIR GmbH, 64291 Darmstadt, Germany}
    \affiliation{GSI Helmholtzzentrum für Schwerionenforschung GmbH, 64291 Darmstadt, Germany}
\author{J. D. Holt}
    \affiliation{TRIUMF, 4004 Wesbrook Mall, Vancouver, BC V6T 2A3, Canada}
    \affiliation{Department of Physics, McGill University, Montréal, Quebec H3A 2T8, Canada}
\author{B. S. Hu}
    \affiliation{Cyclotron Institute and Department of Physics and Astronomy, Texas A\&M University, College Station, Texas 77843, USA}
\author{A. Jaries}
    \affiliation{Max-Planck-Institut für Kernphysik, 69117 Heidelberg, Germany}
\author{C. Klink}
    \affiliation{European Organization for Nuclear Research (CERN), 1211 Geneva 23, Switzerland}
    \affiliation{Institut für Kernphysik, Technische Universität Darmstadt, 64289 Darmstadt, Germany}
\author{Yu. A. Litvinov}
    \affiliation{GSI Helmholtzzentrum für Schwerionenforschung GmbH, 64291 Darmstadt, Germany}
    \affiliation{Institut f\"ur Kernphysik, Universit\"at zu K\"oln, 50937 K\"oln, Germany}
\author{D. Lunney}
    \affiliation{CNRS-TRIUMF International Research Laboratory for Nuclear Physics, Nuclear Astrophysics and Accelerator Technology, CNRS/IN2P3, Vancouver BC, Canada}
\author{V. Manea}
    \affiliation{Université Paris-Saclay, CNRS/IN2P3, IJCLab, 91405 Orsay, France}
\author{F. Mehlhorn}
    \affiliation{Max-Planck-Institut für Kernphysik, 69117 Heidelberg, Germany}
\author{T. Miyagi}
\affiliation{Center for Computational Sciences, University of Tsukuba, 1-1-1 Tennodai, Tsukuba 305-8577, Japan}
\author{M. Mougeot}
\affiliation{Department of Physics, University of Jyväskylä, Accelerator laboratory, FI-40014, Jyväskylä, Finland}
\author{S. Naimi}
    \affiliation{Université Paris-Saclay, CNRS/IN2P3, IJCLab, 91405 Orsay, France}
\author{L. Nies}
    \affiliation{European Organization for Nuclear Research (CERN), 1211 Geneva 23, Switzerland}
    \affiliation{Institut für Physik, Universität Greifswald, 17489 Greifswald, Germany}
\author{M. Schlaich}
    \affiliation{Institut für Kernphysik, Technische Universität Darmstadt, 64289 Darmstadt, Germany}
\author{Ch. Schweiger}
    \affiliation{Max-Planck-Institut für Kernphysik, 69117 Heidelberg, Germany}
\author{L. Schweikhard}
    \affiliation{Institut für Physik, Universität Greifswald, 17489 Greifswald, Germany}
\author{T. Shickele}
    \affiliation{TRIUMF, 4004 Wesbrook Mall, Vancouver, BC V6T 2A3, Canada}
    \affiliation{Department of Physics \& Astronomy, University of British Columbia, Vancouver, BC V6T 1Z1, Canada}
\author{A. Todd}
    \affiliation{TRIUMF, 4004 Wesbrook Mall, Vancouver, BC V6T 2A3, Canada}
    \affiliation{Department of Physics, McGill University, Montréal, Quebec H3A 2T8, Canada}
\author{W. Wojtaczka}
    \affiliation{KU Leuven, Instituut voor Kern- en Stralingsfysica, 3001 Leuven, Belgium}

\begin{abstract}
	\noindent

We present the first determination of the $N = 50$ empirical shell gap at $Z = 48$ by precise mass measurements of the neutron-deficient cadmium isotopes $^{96-98}$Cd with the ISOLTRAP mass spectrometer at ISOLDE-CERN, including the first precise determination of the excitation energy of the $25/2^+$ isomer in $^{97}$Cd. Through the systematics of Coulomb Displacement Energies, we further deduce the empirical shell gap in the higher-$Z$ isotopic chains, 
tightly constraining the $^{100}$Sn mass-surface region.
The new experimental data suggest an enhancement of the gap towards $^{100}$Sn, which is discussed in comparison to state-of-the-art calculations using energy-density functional and new \textit{ab initio} approaches.  

\end{abstract}


\maketitle

The region of the nuclear chart around $^{100}$Sn has been the object of considerable experimental and theoretical efforts in the last years \cite{FAESTERMANN201385,Gorska2022} as it lies at the crossroads of important nuclear-structure features: $^{100}$Sn is the heaviest $N=Z$ doubly magic nucleus, thus nuclear properties measured in this region allow constraining the single-particle degrees of freedom of various models, as well as testing valence-space interactions and \textit{ab initio} theory \cite{Morr18Tin,Miyagi2020,In_99_101_Mougeot,Kart24In,Ge_Ag95,Gust25100Sn} without the computational challenges of mid-shell nuclei. The presence of the $1g_{9/2}$ orbit for both protons and neutrons leads to the formation of many high-spin isomers, which offer unique experimental probes \cite{In_m99_Nies,Maheshwari2025}. Furthermore, the $N \approx Z$ character of these nuclei makes them an ideal testing ground for proton-neutron pairing correlations and isospin symmetry \cite{Cederwall2011}. Finally, the nuclei in the region are 
involved in the rp-process of nucleosynthesis~\cite{Cyburt_2016,Schatz_2017}, making their binding energies important for defining the reaction flow and description of astrophysical observables.  

As outlined in \cite{Gorska2022},
despite remarkable progress by a combination of techniques and advances in sensitivity, there is still a lack of precise experimental information in the region, especially as one approaches $^{100}$Sn, due to production limitations at RIB facilities.
Tin nuclides studied closest to $^{100}$Sn are $^{101}$Sn \cite{101Sn_Ireland} and $^{103}$Sn 
\cite{2023_Xing_CSRe_measurements,Sn103_Nies,Sn103LEBIT},
while for $^{100}$Sn only half-life and $Q_\textrm{EC}$ measurements have been performed so far, the latter having discrepant values \cite{2019_Lubos_fullauthors, Hinke2012_fullauthors}. 
Consequently, the $^{100}$Sn mass determined indirectly from its link to $^{100}$In~\cite{In_99_101_Mougeot}, is of limited precision. 
This problem is compounded by the locally enhanced binding due to the so-called Wigner energy in self-conjugate nuclei~\cite{Wigner} and the crossing of the $N = 50$ closed shell that complicates extrapolations.

Recent mass measurements in the silver isotopic chain have reached $N = 48$ \cite{Ge_Ag95}, allowing a first determination of
the two-neutron, $N = 50$ empirical shell gap for $Z~=~47$. Closer to $^{100}$Sn, however, mass-spectrometry studies have stopped at $N = 50$ for indium ($Z = 49$)~\cite{In_m99_Nies} and cadmium ($Z = 48$) \cite{Giessen_Cd98}, making it impossible to probe the evolution of the gap into the $N \approx Z$ region, or to constrain the proton-neutron interaction in the few-valence-nucleon systems below $N = 50$.

In this work, we present first mass measurements of $^{96,97}$Cd, together with the first determination of the excitation energy of the $J^\pi = 25/2^+$  isomer $^{97n}$Cd.
The $^{96}$Cd measurement was possible due to improved stability of the spectrometer, enabling very long measurements.
We thus determine the $N = 50$ empirical shell gap one step closer to $^{100}$Sn and allow, through local extrapolations,  a first constrained view of its evolution up to $^{100}$Sn. We compare these trends to predictions of self-consistent and of new \textit{ab-initio} calculations within the in-medium similarity renormalization group (IMSRG) framework~\cite{Herg16PR}.

The measurements of neutron-deficient cadmium isotopes reported in this Letter were conducted over two campaigns at the ISOLDE radioactive ion-beam facility at CERN~\cite{Catherall_2017, Borge_2018} using lanthanum carbide (LaC$_\textrm{x}$) targets bombarded by a 1.4-GeV proton beam. 
Reaction products diffused from the heated target into a hot tantalum cavity, where cadmium isotopes were selectively ionized by the resonance ionization laser ion source (RILIS)~\cite{Fedosseev_2017}.
Together with surface-ionized elements, the singly charged cadmium ions were extracted at 30 and 40\,keV and  mass separated through the ISOLDE High Resolution Separator \cite{Catherall_2017}. 
The beams were accumulated and bunched in the radio-frequency quadrupole cooler and buncher (RFQ-cb)~\cite{Herfurth01, KELLERBAUER2001276} of the ISOLTRAP spectrometer \cite{Lunney_2017}. After cooling, the ion bunch was ejected and pulsed down to $3.2\,$keV, for injection into the multi-reflection time-of-flight mass spectrometer (MR-ToF MS)~\cite{WOLF2013123,Wolf_inTrapLift,Wienholtz_Pulse}, where the different species are mass separated by time of flight (ToF) before ejection to an electron-multiplier detector.
The mass $m$ of the ion of interest ($^{96-98}$Cd$^+$) is determined by its ToF $t$, using two reference masses $m_1$ and $m_2$ and their flight times $t_1$ and $t_2$, respectively, using the equation
\mbox{$\sqrt{m} = C_{\textrm{ToF}}\Delta_{\textrm{ref}} + \Sigma_\textrm{ref} / 2$}~\cite{Wienholtz2013}
with \mbox{$\Delta_{\textrm{ref}} = \sqrt{m_1}-\sqrt{m_2}$}, \mbox{$\Sigma_\textrm{ref} = \sqrt{m_1}+\sqrt{m_2}$} and \mbox{$C_{\textrm{ToF}} = (2t-t_1 -t_2)/[2(t_1-t_2)]$} for singly charged ions.
For an isomeric state detected in the same ToF spectrum as its ground state of mass $m_0$ at a ToF $t_0$, the excitation energy can be directly related to the ToF difference $\Delta t$ with \mbox{$E=[(\Delta t /t_0)^2 + 2\Delta t /t_0]m_0c^2$}, where $c$ is the speed of light.

\begin{figure}[t]
	\includegraphics{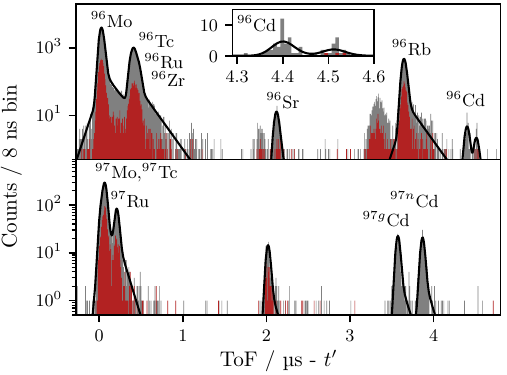}
	\centering
	\caption{ToF spectra of $m/q=96$ (top) and $97$ (bottom) mass-separated beam of a LaC$_\textrm{x}$ target at 1000 revolutions with $t^{\prime} = 23605.8$\,µs and $23728.72$\,µs, respectively. 
    The laser-on data (gray) is modeled using the hyper-EMG PDF~\cite{PURUSHOTHAMAN2017245} (black line). 
    Cadmium ions are unambiguously identified by comparison with ToF data obtained while the RILIS-laser was blocked (red). 
    }
	\label{fig:9697Cd_LaserOnOff}
\end{figure}

The measurements were performed with the ions stored in the MR-ToF MS for 1000 revolutions ($\approx$24\,ms), ensuring detection of the most abundant isobaric contaminant as an in-spectrum reference. 
The observed surface-ionized contaminants are identified by their ToF based on the known mass values of expected isotopes and molecules.
Cadmium ions are unambiguously identified by their correlation to the RILIS-laser presence and wavelength. 
A total of 40 counts of $^{96}$Cd were detected and used for the mass determination. 
During the combined measurement time of 3.5~days, the ionizing laser was 
regularly blocked as a background cross-check, with no $^{96}$Cd counts being detected during this time (see \autoref{fig:9697Cd_LaserOnOff}, top).
The contaminant rate of approximately 400 ions per hour at 1000 revolutions for the coldest target conditions was insufficient for precise ToF-drift corrections. However, for this measurement in addition to stabilizing the mirror voltages ~\cite{Wienholtz_VoltageStab}, we stabilized the MR-ToF MS against temperature variations~\cite{TemperatureStabMRToF}, which reduced ToF-peak drifts below the width of the peak $\Delta t_{\text{FWHM}}$.
With this enhanced stability throughout the entire measurement, it was possible to observe the temporal yield evolution of very close-lying contaminant peaks by analyzing the global composition of the $m/q = 96-98$ MR-TOF MS spectra. 
For instance, the rate of $^{97}$Tc was observed to gradually increase (attributed to in-target feeding from $^{97}$Ru) relative to the rate of the overlapping $^{97}$Mo. Without the temperature stabilization, the two isobars would not have been distinguishable with the obtained mass resolving power \mbox{$R = t/(2\Delta t_{\text{FWHM}})\approx 2\cdot 10^5$}. 

As $^{97n}$Cd partially overlaps in ToF with $^{97}$Rb, the in-trap-decay technique was applied by increasing the RFQ-cb cooling time to approximately three $^{97}$Rb half-lives, which enabled a background-free measurement of $^{97n}$Cd (see \autoref{fig:9697Cd_LaserOnOff}, bottom). The prolonged RFQ-cb trapping also reduced $^{97}$Mo, which formed di-oxide molecules with contaminants in the buffer gas. Therefore, an additional $C_{\textrm{ToF}}$ value for $^{97}$Cd was obtained using $^{97}$Tc as an in-spectrum reference (see \autoref{tab:Results}). 
Where temperature stabilization was not used, the ToF drifts
were corrected using a time-rolling average for the most abundant isobaric contaminant~\cite{FISCHER201844}.
To determine the ground-state $C_{\textrm{ToF}}$ values and the excitation energy of $^{97n}$Cd, simultaneous fits of the in-spectrum reference together with the cadmium isotopes were performed.
The asymmetric ToF peaks were modeled using a multi-component, exponentially modified Gaussian probability density function (hyper-EMG PDF)~\cite{PURUSHOTHAMAN2017245} and an unbinned maximum likelihood estimator~\cite{iminuit, James:1975dr}.

The variance-weighted averages of the $C_{\textrm{ToF}}$ values based on all measurements are presented in \autoref{tab:Results}. The larger of the internal and external errors~\cite{TrappedChargedParticles2008, ErrorCalc} were used as their uncertainties. 
Shifts in $C_{\textrm{ToF}}$ values due to alterations in fit methodology and ToF-drift correction were found to be insignificant.

\autoref{tab:Results} presents the new masses of $^{96-98}$Cd compared to the AME2020 values \cite{AME2020} and \cite{Park_CdAg} for the $25/2^+$ isomer in $^{97}$Cd. 
The mass of $^{98}$Cd is in agreement both with the AME2020 value and the more recent value from \cite{Giessen_Cd98}.
The excitation energy of the $25/2^+$ isomer in $^{97}$Cd is now firmly established at 2246(23)\,keV, following 
the original prediction of $\approx 2.4$\,MeV \cite{Ogawa1983}. This confirmation is important for quantifying the proton-neutron interaction, knowing that $25/2^+$ is the maximum-spin state obtained by coupling the three valence-nucleon holes.

\begin{table*}
\caption{Time-of-flight ratio ($C_\textrm{ToF}$) or difference ($\Delta t_\textrm{ToF}$) and their resulting mass excess and excitation energy of the cadmium isotopes measured in this work. 
The reference masses and literature mass excesses for comparison are from AME2020~\cite{AME2020}, while the literature value for the excitation energy is from Park \textit{et al.}~\cite{Park_CdAg}.
The uncertainties correspond to the statistical uncertainty. 
The half-lives and spin assignments $J^\pi$ are from the NUBASE2020 evaluation~\cite{Kondev_2021}. The value marked with \# is extrapolated from systematics. Order-of-magnitude estimations for the observed yields in the ISOLDE central beamline are  also given. 
}
\begin{ruledtabular}
\begin{tabular*}{\linewidth}{@{\extracolsep{\fill}} cccccccc }
  \multicolumn{6}{c}{}  &\multicolumn{2}{c}{Mass excess or excitation energy (keV)} \\
  \cline{7-8}
 $A$ & $J^\pi$ &Half-life\,(s) &Yield (ions/s)&Reference ions & $C_\textrm{ToF}$ or $\Delta t$\,(ns)  & This Letter  &Literature\\
 \hline
 98   & $0^+$    & 9.29(10) & 100 &$^{98}$Mo$^+$, $^{100}$Cd$^+$ & 
 $0.48904706(570)$ & 
 $-67662(11)$ & 
 $-67640(50)$\\

  \multirow{2}{*}{97}  & \multirow{2}{*}{$9/2^+$}    & \multirow{2}{*}{1.16(5)} & \multirow{2}{*}{1}& $^{97}$Mo$^+$, $^{100}$Cd$^+$ & 
 $0.49029510(893)$ & 
 \multirow{2}{*}{$-60520(24)$} & 
 \multirow{2}{*}{$-60730(420)$}\\

     &    &  & &$^{97}$Tc$^+$, $^{100}$Cd$^+$ & 
 $0.4904272(119)$ & 
 & 
\\
 
97   & $25/2^+$   & 3.86(6) & 1& $^{97g}$Cd$^+$ & 
 $295(3)$ & 
 $2246(23)$ & 
 $2620(580)$\\
 
96   & $0^+$    & 1.003(47) & 0.01 & $^{96}$Mo$^+$, $^{100}$Cd$^+$ &
 0.49105688(815)& 
$-55682(30)$ &
$-55570(410)$\#\\
\end{tabular*}
\label{tab:Results}
\end{ruledtabular}
\end{table*}

The masses determined in this work allow probing the evolution of nuclear structure towards $^{100}$Sn through mass filters, 
such as the one- or two-neutron separation energies $S_{kn} = [M(Z,N-k) + k m_n - M(Z,N)]c^2$, where $k$ is 1 or 2, respectively, $M(Z,N)$ is the atomic mass of the $(Z,N)$ isotope and $m_n$ is the neutron mass. The drop of the separation energy at the crossing of a magic number $N_0$, called the one- or two-neutron empirical shell gap, is determined as $\Delta_{kn}(Z,N_0) = S_{kn}(Z,N_0) - S_{kn}(Z,N_0+k)$. The one-neutron variant is less sensitive to contributions to the binding energy from open-shell correlations, but involves odd nuclei, which are more difficult to compute.

\begin{figure*}[ht]
	\includegraphics[width=0.31\linewidth]{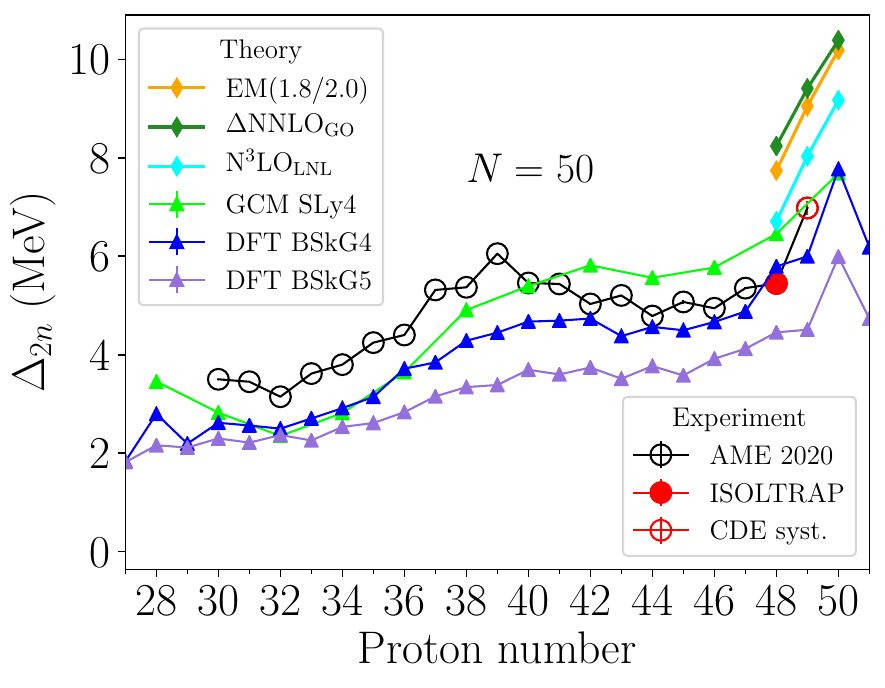}
    \includegraphics[width=0.3\linewidth]{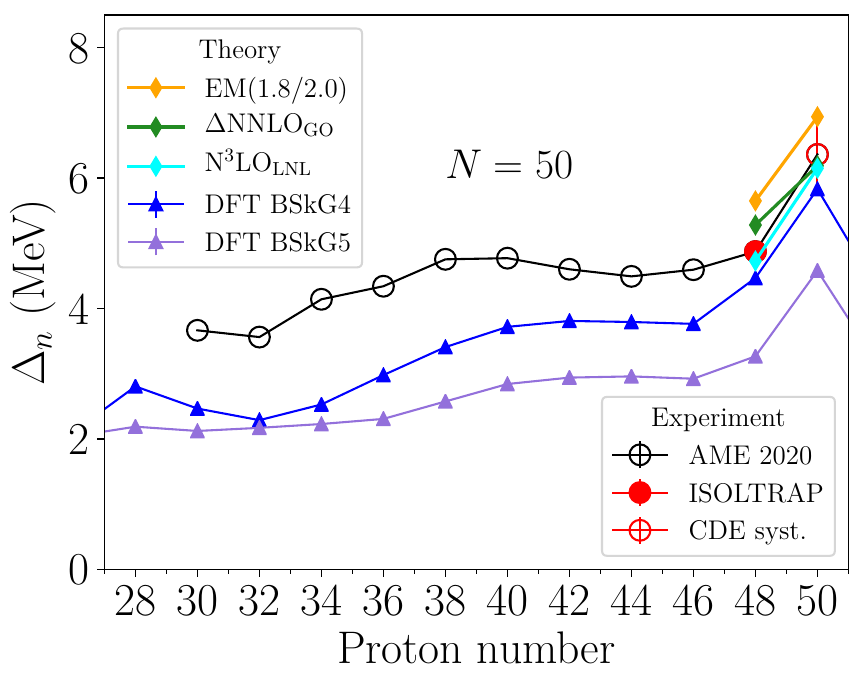}
    \includegraphics[width=0.32\linewidth]{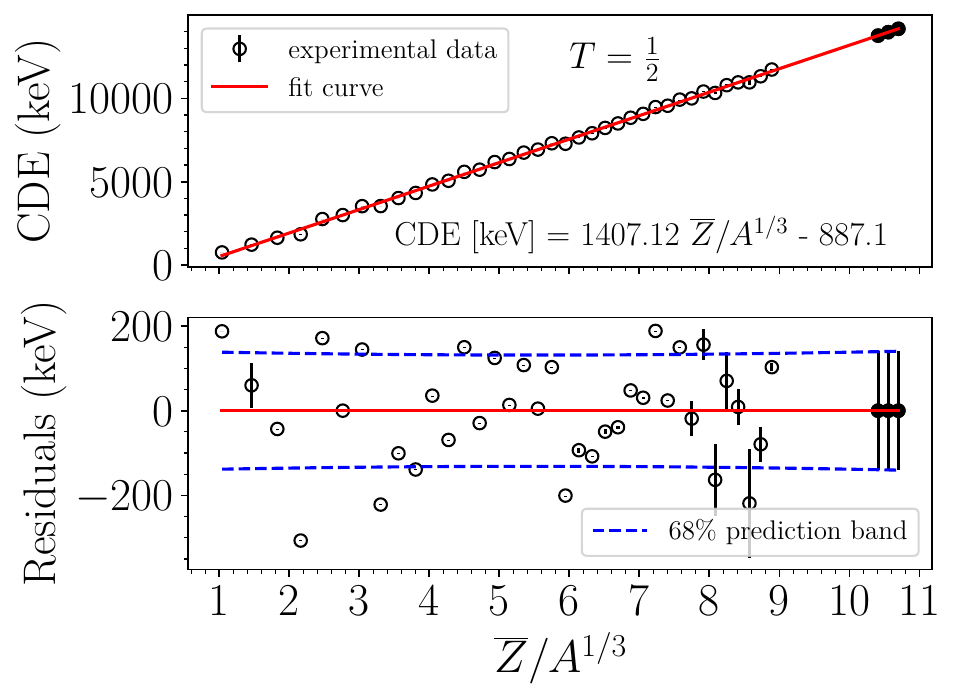}
	\centering
    \caption{$N = 50$ empirical shell gap in its two-neutron (left) and one-neutron (center) variants, compared to theoretical calculations. The experimental data are from the AME2020 (open circles), this work (filled red circles) and extrapolations based on the systematics of CDE (open red circles). The theory data are from the Generator Coordinate Method (GCM) \cite{Bender06} (green triangles) and Density Functional Theory (DFT) \cite{Grams2025,Grams2026} (blue and violet triangles) using Skyrme interactions, as well as new $ab~initio$ calculations (in diamonds) using different chiral interactions in the valence-space in-medium similarity renormalization group (VS-IMSRG): EM(1.8/2.0) \cite{Entem2003,Hebeler2011}  (orange), $\Delta$NNLO$_{\rm GO}$ \cite{Jiang2020}  (dark green) and N$^3$LO$_{\rm LNL}$ \cite{Soma20LNL} (cyan). 
    Right panel shows 
    experimental Coulomb Displacement Energies (CDE) \cite{Antony1997} for the $T=1/2$ isobaric analogues as a function of the parameter $\overline{Z}/A^{1/3}$ and a linear fit. The data are from the AME2020 \cite{AME2020} and \cite{Hockenbery2025} (for $^{75}$Sr). The CDE values extrapolated from the linear trend are marked in filled symbols (and listed in~\autoref{tab:CDE-extrap}).
}
    \label{fig:d2n-dn-50}
\end{figure*}

We determine for the first time the $S_{n}$ and $S_{2n}$  values down to $N = 50$ and thus the corresponding empirical shell gap for $Z = 48$, as shown in \autoref{fig:d2n-dn-50} (center and left, filled red points). The gap is larger than in the lower $Z$ chains, which infers the magic nature of $Z=50$, the crossing of which would likely lead to a local maximum, as in the case of the $N=82$ shell gap \cite{Manea2020}.

Furthermore, we perform well constrained extrapolations, by using the Coulomb Displacement Energy (CDE) between isobaric analogues, expressed by: $CDE = Q_\textrm{EC} + \Delta_{n\textrm{H}}$, where $Q_\textrm{EC}$ is the electron-capture $Q$ value and $\Delta_{n\textrm{H}}$ is the difference between the neutron and hydrogen mass \cite{Antony1997}. For isospin $T = 1/2$, the isobaric analogues are mirror nuclei.
The CDEs show remarkable linearity with the factor $\overline{Z}/A^{1/3}$, where $\overline{Z}$ is the average $Z$ of the pair and $A$ is their mass number.  
The trend of all experimental CDE values for $T = 1/2$ isobaric analogues, based on the AME2020 and \cite{Hockenbery2025} (for $^{75}$Sr) are shown in \autoref{fig:d2n-dn-50} (right), together with 
a linear fit (red). By using the fit parameters (shown on the plot), the CDE values (and thus $Q_\textrm{EC}$) for unknown nuclides can be inferred, allowing to determine their mass. In~\autoref{tab:CDE-extrap}, we present three new masses ($^{97}$In,$^{99}$Sn and $^{95}$Cd) extrapolated  by this method for $T = 1/2$ isobars, based on the $^{97}$Cd mass from this work and recent measurements from 
ISOLTRAP ($^{99}$In) \cite{In_m99_Nies} and JYFLTRAP ($^{95}$Ag) \cite{Ge_Ag95}, respectively. The uncertainties use the $68 \%$ 
prediction interval for the linear regression of the data. 
The new extrapolated values allow to extend $\Delta_{2n}$ to $Z=49$ (\autoref{fig:d2n-dn-50}, left). Combining 
a recent $^{101}$Sn mass from LEBIT \cite{101Sn_Ireland} with the $^{100}$Sn mass determined indirectly in \cite{In_99_101_Mougeot}, we can also derive the $\Delta_{n}$ value for $^{100}$Sn (\autoref{fig:d2n-dn-50}, center).

\begin{table*}
\caption{Mass-excess values of nuclides (``unknown isobars'') deduced using the systematics of Coulomb Displacement Energies~(CDE) and the experimental mass values of their isobaric analogues (``reference nuclides''). The used CDE values from the systematics are given in the third column. AME2020 extrapolations are shown for comparison.}
\begin{ruledtabular}
\begin{tabular}{cccccc}
 Reference nuclide & Mass excess (keV) & CDE (keV) & Unknown isobar & Mass excess (keV) & AME2020 (keV) \\
 \hline
 $^{95}$Ag & -59743.3(1.4) & 13761(140)  &  $^{95}$Cd & -46765(140) & -47060(570)\#\\
 $^{97}$Cd &  -60520(24)& 13966(140) & $^{97}$In & -47336(142) & -47390(400)\# \\
 $^{99}$In &  -61431(14)& 14169(140) & $^{99}$Sn  & -48044(141) & -47980(580)\#\\
\end{tabular}
\label{tab:CDE-extrap}
\end{ruledtabular}
\end{table*}

The values for indium ($\Delta_{2n}$) and tin ($\Delta_{n}$) show a jump with respect to the lower-$Z$ values, marking the effect of the $Z = 50$ magic number, but likely also the enhanced binding of $N = Z$ nuclei, indicating the presence of the Wigner effect in $^{100}$Sn.

In \autoref{fig:d2n-dn-50} we also present recent model calculations. First, we show literature values obtained using the Generator Coordinate Method (GCM) \cite{Bender06} with the SLy4 Skyrme interaction \cite{Chabanat1998} and Density Functional Theory (DFT) calculations using the new BSkG4 \cite{Grams2025} and BSkG5 \cite{Grams2026} Brussels Skyrme interactions, of which the G5 variant contains the first expansion of a Skyrme force to N2LO. The parameters of the SLy4 interaction are fitted using the properties of selected doubly magic nuclei \cite{Chabanat1998}, while the BSkG models use all AME2020 masses. 

The DFT models give a very good qualitative description of the evolution of $\Delta_{n}$ and $\Delta_{2n}$ with $Z$, 
despite a variance in the overall magnitude of the gap, which reflects their difficulty to capture the physics of magic and near-magic nuclei \cite{Grams2025}. The GCM appears better suited near closed shells \cite{Bender06} apparent from the similar agreement despite its more restrained adjustment.  All three models predict the observed enhancement of the empirical gap towards $Z = 50$, noting that the BSkG models also include a phenomenological Wigner term.

We also perform new \textit{ab initio} calculations, using interactions derived from chiral effective field theory~\cite{Epel09RMP,Mach11PR}, constrained by nucleon–nucleon scattering data and selected nuclear properties in few-body, and occasionally heavier, systems, depending on the interaction utilized.
Within the valence-space (VS) formulation of the IMSRG \cite{Stroberg2017,Stroberg2019}, we specifically utilize the EM(1.8/2.0)~\cite{Hebeler2011,Simo17SatFinNuc}, $\Delta$NNLO$_{\rm GO}$ \cite{Jiang2020}, and N$^3$LO$_{\rm LNL}$ \cite{Leis18LNL,Soma20LNL} chiral interactions. 
Through approximate unitary transformations, the VS-IMSRG decouples a chosen valence space, enabling subsequent diagonalization using KSHELL \cite{KSHELL} within the reduced space.

Our calculations start with Hamiltonians in a harmonic-oscillator basis with 13 major shells, and we further restrict three-body matrix elements by imposing $e_1+e_2+e_3 \leq E_{3\rm{max}}\!=\!24$~\cite{Miya22Heavy}. These truncations yield converged binding energies with approximately 2\,MeV uncertainty. For isotopes $N\leq 50$, we choose a $^{56}$Ni core with $0f_{5/2}$, $1p_{1/2}$, $1p_{3/2}$, $0g_{9/2}$ valence orbitals for both protons and neutrons. For isotopes $N\ge50$, a $^{78}$Ni core with $0f_{5/2}$, $1p_{1/2}$, $1p_{3/2}$, $0g_{9/2}$ valence protons and $0g_{7/2}$, $1d_{5/2}$, $1d_{3/2}$, $2s_{1/2}$ and $0h_{11/2}$ valence neutrons is used.
The standard two-body level approximated IMSRG (IMSRG(2)) results tend to show a jump when the valence space is switched.
To mitigate this artifact, we employ the recently developed IMSRG($3\mathrm{f}_{2}$) approximation~\cite{IMSRG3f2}, which extends the IMSRG(2) truncation by incorporating effects of intermediate three-body operators that appear during the valence-space decoupling.

The \textit{ab~initio} calculations show relatively good agreement with the experimental data, especially $\Delta_n$, considering the nuclei with $N\le50$ and $N \ge50$ are computed using different valence spaces.  
These results can be compared to recent lattice chiral EFT calculations \cite{NLEFT} that predict a much larger $\Delta_n$ gap of 12.4\,MeV for $^{100}$Sn but with an uncertainty of 10\,MeV (only 4 isotopes were calculated due to the exascale computing requirements of the lattice approach).  
All calculations predict an enhancement of the empirical gap between $Z = 48$ and $Z = 50$, in agreement with the new experimental data, which is remarkable considering that the interactions are not fit to reproduce magic nuclei, or the Wigner effect, and thus are fully predictive in the region.
This offers promising perspectives for benchmarking  \textit{ab initio} calculations of systematic trends of empirical shell gaps, thus linking the nuclear \textit{ab initio} and mean-field pictures.

In summary, by significantly improving the stability of the ISOLTRAP MR-ToF MS, we have measured the masses of $^{96-98}$Cd, including the $25/2^+$ isomer in $^{97}$Cd and determined for the first time the $N=50$ empirical shell gap for $Z = 48$. By further exploiting the CDEs of $T=1/2$ isobaric analogues, we extended the trend of $\Delta_{2n}$ to $Z=49$ and $\Delta_{n}$ to $Z=50$. We find an enhancement of the empirical shell gap towards $^{100}$Sn, consistent with its doubly magic and self-conjugate nature and in agreement with new predictions of self-consistent and \textit{ab initio} models.
\section*{Acknowledgments} \label{sec:acknowledgements}
The authors gratefully acknowledge the technical support from the ISOLDE operations, target and ion source development and RILIS teams. We also thank the PSB and RP team for the exceptional permission and effort to deliver a proton current above the regular limitation.
We also thank S.~Goriely and W.~Ryssens for making the BSkG5 mass table available to us.

We acknowledge the support of the German Max Planck Society, the French Centre National de Recherche Scientifique (CNRS) and Institut National de Physique Nucl\'{e}aire et de Physique des Particules (IN2P3), 
the European Research Council (ERC) under the European Union’s Horizon 2020 research and innovation programme (Grant Agreements No. 861198 ‘LISA’), as well as the German Federal Ministry of Education and Research (BMBF; Grant No. 05P21HGCI1). 
Furthermore, we acknowledge funding from the European Union's Horizon Europe research and innovation programme under grant agreement no. 101057511.
TRIUMF receives funding via a contribution through the National Research Council of Canada. 
The IMSRG calculations were further supported by NSERC under grants  and SAPIN-2024-00003, the Arthur B. McDonald Canadian Astroparticle Physics Research Institute, the Canadian Institute for Nuclear Physics.
Computations were performed with an allocation of computing resources on Cedar at WestGrid and The Digital Research Alliance of Canada.  
A.B. acknowledges the support of the Natural Sciences and Engineering Research Council of Canada (PDF-587464-2024).
K.B., R.B.C. and Yu.A.L. acknowledge support from the ExtreMe Matter Institute EMMI at the GSI Helmholtzzentrum für Schwerionenforschung GmbH, Darmstadt, Germany.
C.K. acknowledges support from the Wolfgang Gentner Programme of the German Federal Ministry of Education and Research (Grant No. 13E18CHA). 
W.W. acknowledges the support from the Science Research Foundation Flanders (FWO, Belgium), KU Leuven (Grant No. C14/22/104) and FWO Sofina-Boel Fellowship.
T.M. acknowledges the support from the JST ERATO Grant No. JPMJER2304, Japan, and from JSPS KAKENHI Grant Numbers 25K07294, 25K00995, and 25K07330.
\newline
The experiment was conceptualized by M.M.
and conducted by 
D.A., M.Au, M.B., R.B.C., P.F.G., A.H., A.J., C.K., D.La., V.M., F.M., L.N., M.S., Ch.S. and W.W.
\newline
The theoretical calculations were performed by A.B., J.D.H., B.S.H., T.M., T.S. and A.T.
Resources and supervision were provided by K.B., Yu.A.L., D.Lu., S.N. and L.S.
The manuscript was prepared by D.La., D.Lu. and V.M.
All authors contributed to the review and editing of the manuscript.

\section*{Data availability}
The data that support the findings of this article are not publicly available upon publication because it is not technically feasible and/or the cost of preparing, depositing, and hosting the data would be prohibitive within the terms of this research project. The data are available from the authors upon reasonable request.

\bibliographystyle{unsrtnat}
\bibliography{bibliography}

\end{document}